# Evidence of hydrogen diffusion in n-type GaN.

Rafał Jakieła *, Adam Barcz

*Institute of Physics, Polish Academy of Sciences, Aleja Lotnikow 32/46, PL-02668, Warsaw, Poland*

* jakiela@ifpan.edu.pl

The control over impurities like hydrogen and oxygen is of key importance in nitride-based semiconductors due to their unrivaled applicability in optoelectronics and high power/high frequency electronics. Therefore, it is desirable to continue the research on its diffusion and segregation in semiconductor materials. In this work, we report on a first observation on hydrogen outdiffusion from bulk crystalline gallium nitride. The extent of the hydrogen diffusion is established by secondary ion mass spectrometry. Analysis of characteristic hydrogen profile in GaN grown using ammonothermal method, led to the determination of the hydrogen diffusion coefficient at the temperature of 1045ºC - a standard growth temperature for HVPE (halide vapor phase epitaxy) method.

The increase of interest in research on hydrogen in gallium nitride occurred when it was attempted to create a GaN-based p-n junction. Unfortunately, the layers doped with magnesium were highly resistive despite of a very high dopant concentration. Nakamura showed that this is due to the hydrogen passivation of magnesium, which forms inactive complexes with a dopant [1]. The annealing of the layers at suitable temperature results in the breaking of Mg-H bonds and thus activation of magnesium. Interestingly, subsequent experiments with hydrogen in GaN at relatively low temperatures showed a complete lack of hydrogen atom mobility up to 900° C in GaN as well as InN or AlN [2]. It would follow that when heating GaN crystals at high temperatures, hydrogen only breaks bonds with magnesium, but is not removed from the material. Subsequent studies have shown, however, that the acceptor conductivity in the GaN layers grown by MBE can be extinguished by introducing hydrogen into the material [3]. This means that in p-type GaN, hydrogen atoms easily diffuse at 675° C.

Further experiments confirmed that, indeed, H atoms diffuse in p-type material while diffusion in *n*-type one is virtually suppressed. In the very first publications on this subject, deuterium was used as a diffused species. Wilson et.al.[4] studied outdiffusion from D-plasma treated or D-implanted as well as indiffusion into MOMBE-grown semi-insulating GaN. The SIMS depth profiles show deuterium diffusion into SI-GaN to a level of $5\times10^{17}$ cm$^{-3}$ at 250ºC or $10^{18}$ cm$^{-3}$ at 400ºC. Subsequent annealing at 900ºC completely removes hydrogen from GaN layer. The thermal stability of deuterium in n- and p-type D-implanted ($2\times10^{15}$ cm$^{-2}$) GaN was studied by Pearton [5] who observed removal of hydrogen from p-type material at 1000ºC and above with only slight lowering of hydrogen concentration in n-type material even at the temperature of 1200ºC. The Fermi-level dependence of hydrogen diffusion in GaN was also confirmed by Polyakov [6]. Due to the lack of hydrogen diffusion observed in n-type GaN and a problem with hydrogen passivation of GaN:Mg, the further studies were focused on hydrogen diffusion [7, 8, 9] or outdiffusion [10, 11] in p-type material. Our previous study on hydrogenation of GaN:Si layer or GaN:Mg/GaN:Si structure [12] also shows that hydrogen atoms easily diffuse into p-type GaN:Mg from the ambient atmosphere, but no hydrogen diffusion is observed in GaN:Si or through the GaN:Mg/GaN:Si junction. Another question is to, what extent H atoms do not diffuse at all or if the diffusion occurs at the concentration level below hydrogen detection limit. It correlates with the high formation energy and consequently the low solubility of hydrogen in n-type GaN.

Theoretical calculations also predict differences of hydrogen diffusivity [13] as well as interstitial hydrogen formation energy between n- and p-type GaN [13,14,15]. This parameter is ~1.5 eV higher for GaN material with Fermi level at conduction band (n-type material) than at valence band (p-type material) [13,16]. Moreover, hydrogen formation energy in GaN increase as Fermi level moves toward the center of band-gap (material becomes semi-insulating), reaching the highest value ~2 eV at (+/–) transition level [16]. Taking into account the formula for defect concentration in crystalline materials:

$$C_d = N exp\left(\frac{-E_f}{kT}\right) \qquad (1)$$

where: $C_d$ is the defect concentration, $N$ is the number of potential sites in the lattice (per unit volume) where the defect can be created, $E_f$ is the formation energy (energy needed to create a defect), $k$ is the Boltzmann constant, and $T$ is the temperature, hydrogen solubility in semi-insulating GaN under thermodynamic equilibrium does not exceed $10^{15}$ cm$^{-3}$ for octahedral and $2 \times 10^{15}$ cm$^{-3}$ for the tetrahedral position of hydrogen atoms at the temperature of 1000 °C [16]. This amount is far below the detection limit even for such a sensitive method as SIMS. The low formation energy of interstitial hydrogen in p-type GaN (~ -0.1 eV for Fermi level at valence band [16]) allows introduction of high H concentrations at medium temperatures into low p-type material [12]. At the same time, high hydrogen formation energy in n-type GaN (~1 eV for Fermi level at conduction band [16]) allow for hydrogen solubility at the concentration only up to ~$10^{17}$ cm$^{-3}$, even at elevated (over 1000ºC) temperatures. This is the reason why the hydrogen diffusion into n-type or semi-insulating GaN is not observed under standard SIMS measurement conditions, providing the hydrogen detection limit at the level of $10^{17}$ cm$^{-3}$. The different hydrogen diffusion coefficient and the large difference in hydrogen formation energy between p-type and n-type/semi-insulating GaN, do not allow achieving a standard diffusion profile of this species in GaN. This is due to low hydrogen solubility in material with high Fermi level or insufficient thickness of p-type GaN layers – material with low Fermi level. So far, no hydrogen diffusion depth profile in p-type bulk GaN crystal was presented.

In this paper, an alternative approach to hydrogen diffusion in n-type GaN is considered. We report, for the first time, on the observation of hydrogen diffusion in n-type material based on outdiffusion from the bulk crystal. The crystal was grown using ammonothermal method (Am-GaN)[17], with the hydrogen concentration of $3\times10^{18}$ cm$^{-3}$ and n-type carrier concentration of $10^{19}$ cm$^{-3}$ was used as a substrate for the growth of 750 micron-thick layer using HVPE method [18]. The HVPE growth temperature and time were 1045ºC and 4 hours respectively. The growth of GaN layer was preceded by substrate annealing with the following steps:

- from the room temperature up to 600°C – for 30 min, under $N_2$ flow,
- from 600°C up to growth temperature of 1045°C – for 2h, under $N_2$ and $NH_3$ flow
- 1045°C – for 20 minutes, under $N_2$ and $NH_3$ flow

To facilitate the SIMS measurement and keep the proper depth resolution during GaN/GaN interface measurement, the layer was thinned to the 40 microns.

SIMS measurement was performed with a CAMECA IMS6F system using a cesium (Cs+) primary beam, at the energy of 14,5 keV with the current kept at 200 nA. The size of the raster was about 100x100 microns and the secondary ions were collected from a central region of 30 microns in diameter. Secondary ions H¯, Si¯ and Ga¯ as reference signals were detected [19]. The depth profile of silicon was determined for the proper identification of GaN/GaN interface. The H and Si depth profiles in GaN together with *erf* function fitted to hydrogen profile were shown in Fig. 1.

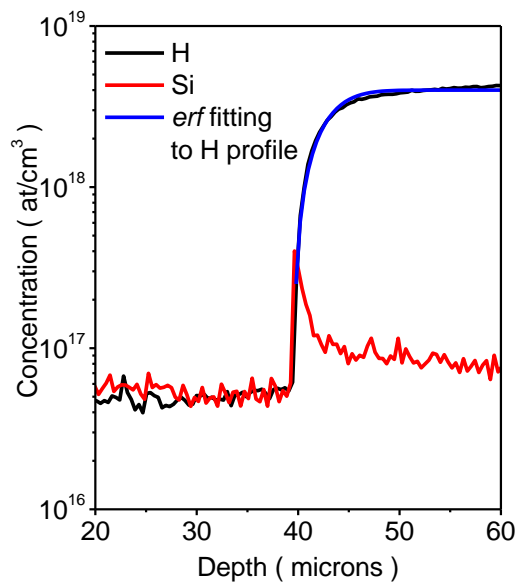

Fig. 1 H and Si depth profiles as well as *erf* function fitted to hydrogen depth profile.

Ammonothermal method brings non-equilibrium conditions for Am-GaN crystal growth in supercritical ammonia ($NH_3$). Such conditions and high hydrogen content in the growth ambient atmosphere provide hydrogen concentration as high as $3\times10^{18}$ cm$^{-3}$ (Fig. 1) even at *n*-type material. Hydrogen-supersaturated Am-GaN crystal subjected to annealing starts to remove surplus hydrogen atoms via out-diffusion. Characteristic discontinuity in hydrogen depth profile (Fig 1.) and lack of hydrogen in HVPE GaN layer with simultaneous hydrogen loss in the Am-GaN substrate (Fig. 2) indicate two phenomena. Firstly, possible hydrogen diffusion into HVPE GaN during growth occurs at a concentration level lower than SIMS detection limit ~$5\times10^{16}$ cm$^{-3}$ (Fig. 1). Secondly, hydrogen outdiffusion from Am-GaN must occur during the substrate annealing before HVPE GaN layer growth.

Since the H diffusion coefficient increases accordingly to the substrate temperature during the annealing, the diffusion coefficient varies with the substrate heating-rate until the final temperature is reached. As a consequence, the *erf* function fitting was conducted using hydrogen depth profile iteration. The relevant iteration is done using Forward Time Center Space (FTCS) method [20], taking into account the activation energy of hydrogen interstitial diffusion in n-type GaN, $E_A = 1.4$ eV, as calculated from *first principles* [21]. This allow to estimate preexponential factor $D_O = 2 \pm 0.5 \times 10^{-6}$ cm$^2$/s. The resultant diffusion coefficients at the temperatures of 900°C and 1045°C amount $1.9 \times 10^{-12}$ and $8.9 \times 10^{-12}$ cm$^2$/s respectively. Simultaneously, the HVPE GaN layer overgrown under equilibrium condition allows incorporation of the out-diffused hydrogen atoms, limited to the specified concentration level determined by hydrogen formation energy in n-type material. Taking into account the background hydrogen concentration (Fig. 1) and equation (1), the lower limit of hydrogen formation energy in GaN amounts 1.55 eV for the undoped HVPE. It indicates on n-type material with Fermi level ~0.5 eV below conduction band [16]. The boundary hydrogen diffusion profiles in undoped n-type HVPE material is shown in Fig. 2

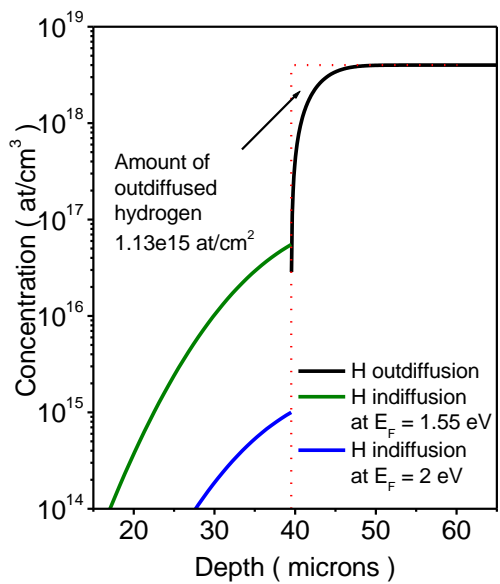

Fig. 2 Error function (*erf*) fitted hydrogen outdiffusion profile from Am-GaN (black line) and simulation of H diffusion profiles into HVPE GaN for hydrogen formation energy 1.55 eV (green line) or 2 eV (blue line).

As the black line represents *erf* fitted hydrogen outdiffusion profile, the area between the red dotted rectangle and H profile characterizes the amount of hydrogen released from Am-GaN during annealing prior to the HVPE GaN layer growth. This area represents a lost hydrogen content of $1.13\times10^{15}$ cm$^{-2}$. Blue and green profiles represent the diffusion profiles within the GaN layer at diffusion coefficient $8.9\times10^{-12}$ cm$^2$/s calculated from *erf* fitted outdiffusion profile and time of GaN layer grown – 4 h, at the temperature of 1045°C. The upper (green) profile is simulated for the diffusion at hydrogen concentration equal to SIMS detection limit, corresponding to the solubility at formation energy of 1.55 eV. The amount of hydrogen diffused into GaN layer at such conditions amounts of ~$2\times10^{13}$ cm$^{-2}$ and is almost two orders of magnitude lower than the dose of the outdiffused species. This difference clearly indicates that the loss of hydrogen from the Am-GaN substrate takes place during annealing prior to the HVPE layer growth. The lower (blue) profile is simulated for diffusion at the calculated diffusion coefficient and formation energy of 2 eV corresponding to the highest $E_F$ value of hydrogen in GaN [16].

Concluding, the experimental findings of this study show for the first time a pure atomic diffusion of hydrogen in *n*-type GaN, described by *erf* function, being the exact solution to the continuity equation. The Fermi-level dependent interstitial hydrogen formation energy in GaN was also shown, based on a lack of measurable depth profile of species in the HVPE grown n-type crystal, resulting from low hydrogen solubility at the growth temperature. Finally, the diffusion coefficient parameters $D_O$ and $E_A$ of hydrogen in n-type GaN were estimated using the previously published calculation and iteratively fitted hydrogen depth profile.